\documentstyle[12pt]{article}
\newcommand{\be}{\begin{equation}}
\newcommand{\ee}{\end{equation}}
\newcommand{\bea}{\begin{eqnarray}}
\newcommand{\eea}{\end{eqnarray}}
\newcommand{\nn}{\nonumber \\}
\newcommand{\p}[1]{(\ref{#1})}
\newcommand{\lb}{\label}

\begin{document}
\begin{titlepage}
\begin{flushright}
{}
\end{flushright}
\vskip 1.3truecm

\begin{center}
{\Large\bf New Model of ${\cal N}{=}8$ Superconformal Mechanics}
\vspace{1.5cm}

{\large\bf F. Delduc$\,{}^a$, E. Ivanov$\,{}^b$,}\\
\vspace{1cm}

{\it a)Laboratoire de Physique de l'ENS Lyon, CNRS UMR5672,}\\
{\it   46, all\'ee d'Italie, 69364 Lyon Cedex 07, France}\\
{\tt francois.delduc@ens-lyon.fr}
\vspace{0.3cm}

{\it b)Bogoliubov  Laboratory of Theoretical Physics, JINR,}\\
{\it 141980 Dubna, Moscow region, Russia} \\
{\tt eivanov@theor.jinr.ru}\\

\end{center}
\vspace{0.3cm}
\vskip 0.6truecm  \nopagebreak

\begin{abstract}
\noindent Using an ${\cal N}{=}4, d{=}1$ superfield approach, we construct
an ${\cal N}{=}8$ supersymmetric
action of the self-interacting off-shell ${\cal N}{=}8$ multiplet ${\bf (1, 8, 7)}$. This
action is found to be invariant under the exceptional ${\cal N}{=}8, d{=}1$
superconformal group $F(4)$ with
the $R$-symmetry subgroup $SO(7)$.
The general ${\cal N}{=}8$ supersymmetric ${\bf (1, 8, 7)}$ action is a sum
of the superconformal action
and the previously known free bilinear action. We show that the general action
is also superconformal, but with respect to redefined superfield transformation
laws. The scalar potential
can be generated by two Fayet-Iliopoulos ${\cal N}{=}4$ superfield terms
which preserve
${\cal N}{=}8$ supersymmetry but break the superconformal and $SO(7)$ symmetries.
\end{abstract}
\vspace{1.1cm}

\newpage

\end{titlepage}
\section{Introduction}
Superconformal mechanics (SCM) plays an important role in a wide circle of intertwining
problems related to black holes, AdS$_2$/CFT$_1\,$, super Calogero-Moser models,
branes, etc \cite{AP}-\cite{BIKL}. While the ${\cal N}{=}2$ and
${\cal N}{=}4$ SCM models were constructed
long ago \cite{AP}-\cite{AGK}, much less is known about the higher ${\cal N}{>}4$ cases,
in particular, the ${\cal N}{=}8$ one. This is related to the fact that the number
of admissible non-equivalent $d{=}1$
superconformal groups is growing with ${\cal N}$. For instance, there is only one choice for
the ${\cal N}{=}2$ superconformal group, $SU(1,1\vert 1) \sim OSp(2\vert 2)$,
whereas there are three possibilities, $D(2,1; \alpha)$,
$OSp(4^\star\vert 2)$ and $SU(1,1\vert 2)$,
in the ${\cal N}{=}4$ case\footnote{Actually, the supergroups $OSp(4^\star\vert 2)$
and $SU(1,1\vert 2)$ can be treated as special cases of
$D(2,1; \alpha)\,$.}
and four in the ${\cal N}{=}8$ case: $OSp(8\vert 2)\,$,
$OSp(4^\star\vert 4)\,$, $F(4)$ and $SU(1,1\vert 4)$ \cite{VP,FRS}.

The ${\cal N}{=}8$ superconformal actions of the off-shell multiplets ${\bf (5, 8, 3)}$
and ${\bf (3, 8, 5)}$
were explicitly given in \cite{BIKL} in terms of the properly constrained ${\cal N}{=}4$, $d{=}1$
superfields. It was found that in both cases the underlying superconformal symmetry
is $OSp(4^\star\vert 4)\,$.
It is interesting to construct superconformal models associated with other
${\cal N}{=}8, d{=}1$ superconformal groups. An example of such a system is presented here.
It is the ${\cal N}{=}8$ supersymmetric mechanics model associated with
the off-shell multiplet ${\bf (1, 8, 7)}$ from the list of
\cite{ABC}. The underlying ${\cal N}{=}8$ superconformal symmetry is the exceptional
supergroup $F(4)$ with the $R$-symmetry subgroup $SO(7)$. In
the ${\cal N}{=}4$ superfield approach
which we use throughout the paper, the ``manifest'' superconformal group is
$D(2,1; -1/3) \subset F(4)\,$.

In terms of ${\cal N}{=}4$ superfields, the multiplet in question amounts to a sum
\be
{\bf (1, 8, 7)} = {\bf (1, 4, 3)} \oplus {\bf (0, 4, 4)}. \label{8_4}
\ee
While for the multiplet $ {\bf (1, 4, 3)}$ one can write manifestly ${\cal N}{=}4$
supersymmetric
actions in the ordinary ${\cal N}{=}4$ superspace \cite{leva,ikp}, actions
of the fermionic ${\cal N}{=}4$ multiplet
$ {\bf (0, 4, 4)}$ are naturally written in the analytic harmonic
${\cal N}{=}4$ superspace \cite{IL,BIKL,ABC}.
In order to construct the ${\cal N}{=}8$ supersymmetric actions of
the multiplet $ {\bf (1, 8, 7)}$
we use the ${\cal N}{=}4, d{=}1$ harmonic superspace description for
both ${\cal N}{=}4$
multiplets in \p{8_4}.
A sum of the free ${\bf (1, 8, 7)}$ action \cite{ABC}
and the superconformal action constructed here yields the most general
$ {\bf (1, 8, 7)}\,$ action.
We present the relevant component off-shell action and show that it agrees with that found
in \cite{Topp} by a different method. Surprisingly, the general action is also superconformal,
though with respect to redefined superfield transformation laws.
For both ${\cal N}{=}4$ multiplets one can construct ${\cal N}{=}8$ supersymmetric
Fayet-Iliopoulos terms which, however, break superconformal symmetry.

\section{Preliminaries: ${\cal N}{=}4$, $d{=}1$ harmonic superspace}
The harmonic analytic ${\cal N}{=}4$ superspace \cite{HSS,IL,1,2} is parametrized
by the coordinates
\be
(\zeta, u) = (t_A, \theta^+, \bar\theta^+, u^\pm_i)\,, \quad u^{+ i}u_i^- =1\,.
\ee
They are related to the standard ${\cal N}{=}4$ superspace (central basis) coordinates
$z = ( t, \theta_i, \bar\theta^i)$ as
\be
t_A = t -i (\theta^+\bar\theta^- + \theta^-\bar\theta^+), \quad \theta^\pm
= \theta^iu^\pm_i\,, \;
 \bar\theta^\pm = \bar\theta^iu^\pm_i\,.
\ee
The ${\cal N}{=}4$ covariant spinor derivatives and their harmonic projections
are defined by
\bea
&& D^i = \frac{\partial}{\partial \theta_i} + i\bar\theta^i \partial_t\,, \;\;
\bar D_i = \frac{\partial}{\partial \bar\theta^i} + i\theta_i \partial_t\,,
\;\; \overline{(D^i)}
= -\bar D_i\,, \;\;\{D^i, \bar D_k \} = 2i\,\delta^i_k\partial_t\,,
\label{defD2} \\
&& D^\pm = u^\pm_i D^i\,,\quad \bar D^\pm = u^\pm_i \bar D^i\,, \quad
 \; \{D^+, \bar D^- \} = - \{D^-, \bar D^+ \}
= 2i\,\partial_{t_A}\,. \label{defD1}
\eea
In the analytic basis, the derivatives $D^{+}$ and $\bar D^+$ are short,
\be
D^+ = \frac{\partial}{\partial \theta{}^-}\,, \quad \bar D^+ =
-\frac{\partial}{\partial \bar\theta{}^-}\,.
\ee
The analyticity-preserving harmonic derivative $D^{++}$ and its conjugate
$D^{--}$ are given by
\bea
&& D^{++}=\partial^{++}-2i\theta^+\bar\theta^+\partial_{t_{A}}
+\theta^+\frac{\partial}{\partial\theta^-}
+ \bar\theta^+\frac{\partial}{\partial\bar\theta^-}\,, \nn
&& D^{--}=\partial^{--}-2i\theta^-\bar\theta^-\partial_{t_{A}}
+\theta^-\frac{\partial}{\partial\theta^+}
+  \bar\theta^-\frac{\partial}{\partial\bar\theta^+}\,, \quad
\partial^{\pm\pm} = u^{\pm}_i\frac{\partial}{\partial u^{\mp}_i}\,,
\eea
and become the pure partial derivatives $\partial^{\pm\pm}$ in the central basis.
They satisfy the relations
\be
[D^{++},D^{--}]= D^{0}\,, \quad [D^0, D^{\pm\pm}] =
\pm 2 D^{\pm\pm}\,, \lb{DharmAl}\\
\ee
where $D^0$ is the operator counting external harmonic $U(1)$ charges.
The integration measures in the full harmonic superspace (HSS) and
its analytic subspace are defined as
\begin{eqnarray}
&& \mu_H = dudtd^4\theta=dudt_{A}(D^-\bar D^-)(D^+\bar D^+)
=\mu_{A}^{(-2)}(D^+\bar D^+)\,,\nn
&& \mu_{A}^{(-2)}=dud\zeta^{(-2)}
=dudt_{A}d\theta^+d\bar\theta^+=dudt_{A}(D^-\bar D^-)\,. \label{measures}
\end{eqnarray}

\setcounter{equation}{0}
\section{The multiplets (1, 4, 3) and (0, 4, 4)}

\subsection{(1, 4, 3)}
The off-shell multiplet ${\bf (1, 4, 3)}$ is described by
a real ${\cal N}{=}4$  superfield $v(z)$ obeying the constraints \cite{leva}
\be
D^iD_i v = \bar D_i\bar D^i v = 0\,, \quad [D^i, \bar D_i] v = 0\,. \label{Uconstr1}
\ee
The same constraints in HSS read \cite{ABC}
\be
D^{++}v =0\,, \;\; D^+D^-v = \bar D^+\bar D^-v = 0\,, \quad
\left(D^+\bar D^- + \bar D^+D^-\right)v = 0\,.\label{Uconstr2}
\ee
The extra harmonic constraint guarantees the harmonic independence of
$v$ in the central basis.

Recently, it was shown \cite{2} that this multiplet can be also described in terms
of the real analytic gauge superfield ${\cal V}(\zeta, u)$ subjected to
the abelian gauge transformation
\be
{\cal V} \;\;\Rightarrow \;\; {\cal V}{\,}' = {\cal V} + D^{++}\Lambda^{--}\,, \quad
\Lambda^{--} = \Lambda^{--}(\zeta, u)\,.\label{VgaugeT}
\ee
In the Wess-Zumino gauge just the irreducible ${\bf (1, 4, 3)}$ content remains
\be
{\cal V}_{WZ}(\zeta, u) = x(t_A) + \theta^+\omega^i(t_A) u^-_i
+ \bar\theta^+ \bar\omega^i(t_A) u^-_i +
3i \theta^+\bar\theta^+ A^{(ik)}(t_A)u^-_iu^-_k\,. \label{WZV1}
\ee
No residual gauge freedom is left. The original superfield $v(z)$ is
related to ${\cal V}(\zeta, u)$ by
\be
v(t, \theta^i, \bar\theta_k)=
\int du\, {\cal V}\left(t -2i\theta^i\bar\theta^k u^+_{(i}u^-_{k)}\,,\, \theta^iu^+_i,
\bar\theta^ku^+_k\,,\, u^\pm_l\right). \label{DefU2}
\ee
The constraints \p{Uconstr1} are recovered as a consequence of the harmonic
analyticity of  ${\cal V}$
\be
D^+{\cal V} = \bar D^+{\cal V} = 0\,.\label{Vanalit}
\ee

We shall need a ``bridge'' representation of ${\cal V}$
through the superfields $v(z)$ and $V^{--}(z, u)$
\be
{\cal V} = v + D^{++} V^{--}\,, \quad v {\,}' = v \,, \;\; V^{--}{\,}' =
V^{--} + \Lambda^{--}\,.
\label{Br}
\ee
The term $v(z)$ is just given by the expression \p{DefU2}.
The analyticity conditions \p{Vanalit} imply
\bea
D^-v + D^{+}V^{--} = \bar D^-v + \bar D^{+}V^{--} = 0\,. \label{Dconstr}
\eea
Below are some useful corollaries of \p{Dconstr}, \p{Vanalit} and \p{defD1}
\bea
&& \left(D^+\bar D^- - \bar D^+ D^-\right)v = -2 D^+\bar D^+\,V^{--}\,, \label{1} \\
&& D^+\bar D^+\,V^{--} = i D^{++}\left(\dot{V}{}^{--}+\frac{i}{2}\,D^-\bar D^-v\right)
- i \dot{{\cal V}} \,,
\label{2} \\
&& D^+\left(\dot{V}{}^{--} + \frac{i}{2}\,D^-\bar D^-v\right) =
\bar D^+\left(\dot{V}{}^{--} + \frac{i}{2}\,D^-\bar D^-v\right) = 0\,. \label{3}
\eea

The general invariant action of the multiplet ${\bf (1, 4, 3)}$ reads
\be
S_{gen}^{(v)} = \int dtd^4\theta\, {\cal L}_{gen}(v)\,.\label{genact_v}
\ee
The free action corresponds to the quadratic Lagrangian
\be
S_{free}^{(v)} = -\int dtd^4\theta\,v^2\,, \label{free_v}
\ee
while the action invariant under the most general ${\cal N}{=}4, d{=}1$
superconformal group
$D(2,1; \alpha)$ (except for the special values of $\alpha{=}0,-1$)
is \cite{ikl1}
\be
S_{sc}^{(v)} =  -\int dtd^4\theta\,(v)^{-\frac{1}{\alpha}}\,, \label{conf_v}
\ee
where, for the correct $d{=}1$ field theory interpretation, one must assume
that $v$
develops a non-zero background value,
$v = 1 + \ldots \,$. The transformation properties of some
relevant objects under
the conformal ${\cal N}{=}4$ supersymmetry $\subset D(2,1; \alpha)$
are as follows \cite{IL,2}\footnote{Invariance under these transformations
is sufficient to check
$D(2,1; \alpha)$ invariance since the rest of
the $D(2,1; \alpha)$
transformations is contained in the closure of the conformal and
manifest Poincar\'e ${\cal N}{=}4, d{=}1$ supersymmetries.}
\bea
&& \delta_{sc}\, D^{++} = -\Lambda^{++}_{sc}\, D^0\,,
\quad \delta_{sc}\, D^0 = 0\,, \label{DDConf} \\
&& \delta_{sc}\, \mu_A^{(-2)} = 0\, \quad \delta_{sc}\, \mu_H =
 \mu_H\left(2\Lambda_{sc}
 - \frac{1 + \alpha}{\alpha}\,\Lambda_0 \right)\,, \nn
&& \delta_{sc}\,(dt d^4\theta)
= -\frac{1}{\alpha}\,(dt d^4\theta)\,\Lambda_0\,, \quad
\delta_{sc}\,du = du\,D^{--}\Lambda^{++}_{sc}\, \label{measConf} \\
&& \delta_{sc}\, v = -\Lambda_0\,v\,, \quad \delta_{sc}\,{\cal V} =
-2\Lambda_{sc}\,{\cal V}\,,\label{vVconf}
\eea
where
\bea
&& \Lambda^{++}_{sc} =
2i\alpha(\bar\varepsilon^+ \theta^+ {-} \varepsilon^+ \bar\theta^+ )
\equiv D^{++}\Lambda_{sc}\,, \, \Lambda_{sc} =
2i\alpha(\bar\varepsilon^- \theta^+ {-} \varepsilon^- \bar\theta^+ )\,, \,
(D^{++})^2\Lambda_{sc} = 0\,,
\label{Conf1} \\
&&\Lambda_0 = \left(2\Lambda_{sc} - D^{--}\Lambda^{++}_{sc}\right) =
2i\alpha\left(\theta_i\bar\varepsilon^i + \bar\theta^i\varepsilon_i\right),
\quad D^{++}\Lambda_0 = 0\,, \label{Conf2}
\eea
and $\varepsilon^{\pm} = \varepsilon^iu^\pm_i\,, \;\bar\varepsilon^{\pm} =
\bar\varepsilon^iu^\pm_i\,$, $\varepsilon^i, \bar\varepsilon_i$
being mutually conjugated Grassmann transformation parameters.
Using \p{measConf}, \p{vVconf}
and \p{Conf1}, \p{Conf2}, it is easy to check
the $D(2,1; \alpha)$ invariance
of the action \p{conf_v} and the covariance of the relation \p{DefU2}.

One can also construct an ${\cal N}{=}4$ supersymmetric Fayet-Iliopoulos (FI) term
\be
S_{FI}^{(v)} = i\int du d\zeta^{(-2)}\, c^{+2}\,
{\cal V}\, \quad c^{+2} = c^{ik}u^+_iu^+_k\,, \;[c] = cm^{-1}\,,\label{FI1}
\ee
which produces a scalar potential after elimination of the auxiliary field
$A^{(ik)}$ in the sum of
\p{genact_v} and \p{FI1}. This term is superconformal only for the special choice
$\alpha{=}0$ \cite{2}.

\subsection{(0, 4, 4)}
The multiplet ${\bf (0,4,4)}$ comprises 4 fermionic fields and 4 bosonic
auxiliary fields. It is described off shell by the fermionic
analytic superfield $\Psi^{+ A}$,
$\widetilde{(\Psi^{+ A})} = \Psi_A^+\,$, obeying the constraint \cite{IL}:
\be
D^{++}\Psi^{+ A} = 0\,\; \Rightarrow \; \Psi^{+ A} =
\psi^{iA}u^+_i + \theta^+ a^A + \bar\theta^+ \bar{a}^A
+ 2i\theta^+\bar\theta^+ \dot{\psi}{}^{i A}u^-_i\,.  \label{PsiConstr}
\ee
With respect to the doublet index $A$ ($A = 1,2$), it is transformed
by some extra (``Pauli-G\"ursey'')
group $SU(2)_{PG}$ which commutes with ${\cal N}{=}4$ supersymmetry.
The requirement of superconformal
covariance of the constraint \p{PsiConstr} uniquely fixes the superconformal
$D(2,1;\alpha)$ transformation rule of $\Psi^{+ A}$, for any $\alpha$, as
\be
\delta_{sc}\,\Psi^{+ A} = \Lambda_{sc}\, \Psi^{+ A}\,.\label{traNPsi1}
\ee
In the central basis, the constraint \p{PsiConstr} implies
\be
\Psi^{+ A}(z, u) = \Psi^{i A}(z)u^+_i\,,  \label{CBpsi}
\ee
and the analyticity conditions $D^+\Psi^{+ A} = \bar D^+\Psi^{+ A} =0$ amount to
\be
D^{(i} \Psi^{k) A}(z) = \bar D^{(i} \Psi^{k) A}(z) = 0\,. \label{ConstrPsi2}
\ee

The free action of $\Psi^{+ A}$,
\be
S^{(\psi)}_{free} = \int du d\zeta^{(-2)}\,\Psi^{+ A}\Psi^{+}_{A}\,,
\label{Freepsi}
\ee
is not invariant under $D(2,1;\alpha)$ (except for the special case of
$\alpha{=}0$). However,
we can construct a superconformal invariant by coupling $\Psi^{+ A}$
to the ${\bf (1,4,3)}$ multiplet \cite{2}
\be
S^{(\psi)}_{(sc)} =\int du d\zeta^{(-2)}\,{\cal V}\, \Psi^{+ A}\Psi^{+}_{ A}\,.
\label{Confpsi}
\ee
This action is superconformal at any $\alpha $, and it also respects
the gauge invariance
\p{VgaugeT} as a consequence of the constraint \p{PsiConstr}. Assuming that
${\cal V} = 1 + \widetilde{{\cal V}}$, $v = 1 + \widetilde{v}$, \p{Confpsi} can be
treated as a superconformal generalization of the free action \p{Freepsi}.
A simple analysis
based on dimensionality and on the Grassmann character of the superfields
$\Psi^{+ A}, \Psi^{- A} = D^{--}\Psi^{+ A}$ shows that no self-interaction
of the multiplet ${\bf (0,4,4)}$ can be constructed. Also, the coupling
\p{Confpsi} is
the only possible coupling of this fermionic multiplet to the
multiplet ${\bf (1,4,3)}$
preserving the canonical number of fields with time derivative
in the component action
(no more than two for bosons and no more than one for fermions).

The only additional ${\cal N}{=}4$ invariant one can construct
is the appropriate FI-type term
\be
S^{(\psi)}_{FI} = \int du d\zeta^{(-2)} \left( \theta^+ \xi_A\Psi^{+ A}
+ \bar\theta^+ \bar\xi^A\Psi^+_A \right), \label{FIpsi}
\ee
$\xi_A, \bar\xi^A$ being $SU(2)_{PG}$ breaking constants.
This term is superconformal at $\alpha{=}-1$ \cite{2}.

\setcounter{equation}{0}
\section{${\cal N}{=}8$ supersymmetry}
As shown in \cite{ABC},
one can define the hidden ${\cal N}{=}4$ supersymmetry\footnote{The
relative sign between these two
variations was chosen so as to have the same closure for the hidden
supersymmetry as for the manifest
one.}
\be
\delta_\eta \,v = -\eta_{iA}\Psi^{iA}\,, \qquad \delta_{\eta}\,\Psi^{iA} =
\frac{1}{2}\,\eta^A_k\,\left( D^{i}\bar D^{k} - \bar D^{i} D^k \right) v\,. \label{N81}
\ee
It commutes with the explicit ${\cal N}{=}4$ supersymmetry and so forms, together
with the latter, ${\cal N}{=}8,\; d{=}1$ Poincar\'e supersymmetry. It is easy
to check the compatibility
of \p{N81} with the constraints \p{Uconstr1}, \p{ConstrPsi2}. The same
transformations,
being rewritten in HSS, read
\bea
\delta_\eta\, v = \eta^{- A}\Psi^+_A - \eta^{+ A}\Psi^-_A\,, \;
\delta_\eta\, \Psi^{+ A} = \eta^{- A}\,D^+\bar D^{+}v - \frac{1}{2}\eta^{+ A}
\left( D^+\bar D^{-} - \bar D^+ D^{-}\right)v\,, \label{N82}
\eea
where $\eta^{\pm A} = \eta^{i A}u^\pm_i\,$. The appropriate transformation
of the analytic prepotential
${\cal V}$ is
\be
\delta_\eta {\cal V}(\zeta, u)  = 2 \eta^{- A}\Psi^{+}_{A}(\zeta, u)\,. \label{N8V}
\ee
As expected, \p{N8V} closes on the time derivative of ${\cal V}$
only modulo a gauge transformation:
\bea
[\delta_\eta, \delta_{\eta{\,}'}]{\cal V} = 2i (\eta{\,}'_{i A}\eta^{iA})
\left(\dot{{\cal V}}
- D^{++}\widetilde{\Lambda}{}^{--} \right), \quad
\widetilde{\Lambda}{}^{--} =\dot{V}{}^{--}
+ \frac{i}{2}\,D^-\bar D^{-} v\,,\label{CommV}
\eea
where we used the identities \p{1}-\p{3} and anticommutation
relations \p{defD1}.

Now we wish to construct the most general action of the multiplets ${\bf (1,4,3)}$
and ${\bf (0,4,4)}$ which would enjoy the hidden supersymmetry \p{N81}, \p{N82}
and so present the ${\cal N}{=}4$ superfield form of the general ${\cal N}{=}8$
supersymmetric action of the multiplet ${\bf (1,8,7)}\,$.

A convenient starting point of such a construction is offered by
the $\Psi$-actions \p{Freepsi}
and \p{Confpsi} in view of their uniqueness. The ${\cal N}{=}8$
completion of
the free action \p{Freepsi} was found in \cite{ABC}. The variation
of \p{Freepsi} under \p{N82} can be written as
\be
\delta_\eta \,S^{(\psi)}_{free} =
4 \int \mu^{(-2)}_A\, D^+\bar D^+\,v\, (\eta^{-A}\Psi^+_A) =
4 \int \mu_H\,v (\eta^{-A}\Psi^+_A)
= -2\int dt d^4\theta\, v(\eta^{iA}\Psi_{iA})\,, \label{VarS1}
\ee
where we used the relation \p{measures}, constraint \p{PsiConstr}
and the harmonic independence
of $v$ in the central basis. This variation is cancelled by that of
the free $v$ action \p{free_v}, so the action
\be
S_{free}^{({\cal N}{=}8)} =
\frac{1}{2}\left(\int\mu^{(-2)}_A\,\Psi^{+ A} \Psi^+_A
-\int dtd^4\theta\,v^2\right)  \label{freeN8}
\ee
is ${\cal N}{=}8$ supersymmetric. It breaks superconformal symmetry,
since its first term
is invariant under $D(2,1; \alpha{=}-1/2)$ (see \p{conf_v}),
while the second one is invariant under
$D(2,1; \alpha{=}0)\,$.

Now let us promote the interaction action \p{Confpsi} to an ${\cal N}{=}8$ invariant.
To calculate the variation $\delta_\eta\, S^{(\psi)}_{sc}$,
we firstly note that it is fully specified by the variation $\delta_\eta \Psi^{+ A}$,
since $\delta_\eta\,{\cal V}\,$ does not contribute because of the nilpotency
property $(\Psi^{+ A})^3 = 0\,$. Then, using \p{1}, we represent
\be
\delta_\eta \Psi^{+ A} = D^+\bar D^{+}\left(\eta^{- A}v
+ \eta^{+ A} V^{--}\right),
\ee
restore the full superspace integration measure in $\delta_\eta\, S^{(\psi)}_{sc}$
and rewrite
this variation as
\bea
\delta_\eta\, S^{(\psi)}_{sc}
&=& 2\int \mu_H \left[v^2\,(\eta^{-A}\Psi^+_A)
+ v D^{++}V^{--} (\eta^{-A}\Psi^+_A) \right. \nn
&& \left. +\; v V^{--}(\eta^{+A}\Psi^+_A)
+ V^{--}D^{++}V^{--} (\eta^{+A}\Psi^+_A)\right],\label{VarS2}
\eea
where the bridge representation \p{Br} for ${\cal V}\,$ was used. Taking into account the
harmonic constraint \p{PsiConstr} and the properties $D^{++}\eta^{- A} = \eta^{+ A}$ and
$D^{++}\eta^{+ A} = 0\,$, we observe that all terms in \p{VarS2}
except for the first one are
reduced to a total harmonic derivative, whence
\be
\delta_\eta\, S^{(\psi)}_{sc} = 2 \int \mu_H \,v^2\,(\eta^{-A}\Psi^+_A) =
- \int dtd^4\theta (\eta^{iA}\Psi_{iA}) v^2\,.\label{VarS3}
\ee
This is cancelled out by the variation of $-1/3 \int dt d^{4}\theta\,v^3\,$, so
the second ${\cal N}{=}8$ supersymmetric action is given by
\be
S_{sc}^{({\cal N}{=}8)} = \frac{1}{2}\left(\int\mu^{(-2)}_A\,{\cal V}\,\Psi^{+ A} \Psi^+_A
- \frac{1}{3}\int dtd^4\theta\,v^3\right).\label{confN8}
\ee

Since the first term in \p{confN8} is $D(2,1; \alpha)$ invariant at any $\alpha $, while
the second one is invariant under $D(2,1;\alpha{=}-1/3)$ (see eq. \p{conf_v}), we conclude
that the full action \p{confN8} is invariant under the ${\cal N}{=}4$ superconformal group
$D(2,1;\alpha{=}-1/3)\,$. Since it is also invariant under the rigid
${\cal N}{=}8, d{=}1$ supersymmetry,
it is invariant under some ${\cal N}{=}8$ superconformal group.
Therefore the action \p{confN8},
provided that ${\cal V}$ and $v$ start with a constant, ${\cal V} = 1 + \tilde{{\cal V}}\,$,
$v = 1 + \tilde{v}\,$, defines a new model of ${\cal N}{=}8$ superconformal mechanics
associated
with the ${\cal N}{=}8$ multiplet ${\bf (1, 8, 7)}\,$. It is easy
to recognize which ${\cal N}{=}8, d{=}1$ superconformal group we are facing
in the present case. As follows from \cite{FRS}, the only such supergroup
in which one can embed $D(2,1;\alpha{=}-1/3)\,$ (to be more exact, an equivalent
supergroup $D(2,1;\beta)\,$, $\beta{=}-\frac{1 +\alpha}{\alpha}{=}2$) is
the exceptional ${\cal N}{=}8, d{=}1$ superconformal group $F(4)\,$, with the
$R$-symmetry subgroup $SO(7)$.

As already mentioned, the actions \p{Freepsi} and \p{Confpsi} are the unique
$d{=}1$ sigma model type actions simultaneously involving both
the ${\bf (0, 4, 4)}\,$ and
${\bf (1, 4, 3)}\,$ multiplets. Hence the sum of
their ${\cal N}{=}8$ completions, \p{freeN8} and \p{confN8},
yields the most general ${\cal N}{=}8$ supersymmetric sigma-model
type off-shell action of the multiplet
${\bf (1, 8, 7)}\,$ in the ${\cal N}{=}4$ superfield formulation:
\be
S_{gen}^{({\cal N}{=}8)} = \gamma\,S_{free}^{({\cal N}{=}8)}
+ \gamma{\,}'\,S_{sc}^{({\cal N}{=}8)}\,, \label{genN8}
\ee
$\gamma$ and $\gamma{\,}'$ being two independent renormalization constants.
Surprisingly,
it is also $F(4)$ invariant, though with respect to modified
superfield transformation
laws. Choosing, for simplicity, $\gamma = 1$, and making the redefinitions
$ 1 + \gamma{\,}'\, {\cal V} = \tilde{{\cal V}}\,$, $1 + \gamma{\,}'\,v
= \tilde{v}\,$, $\gamma{\,}' \Psi^{+ A} = \tilde{\Psi}{}^{+ A}\,$,
we observe that \p{genN8} is reduced, up to a constant renormalization factor,
to \p{confN8} where all superfields are replaced by those with ``tilde''.
Assuming for the
new superfields the same transformation laws \p{vVconf}, \p{traNPsi1}
as for the original ones, we see
that \p{genN8} is also $F(4)$ invariant. The $F(4)$ transformations of the original
superfields $v$ and ${\cal V}$ are of course modified, e.g. $\delta'_{sc}v =
-\Lambda_0\,[v  +(\gamma{\,}')^{-1}]\,$. The transformations of
the hidden ${\cal N}{=}4$
supersymmetry remain unaltered. There is no way to make superconformal
the free action \p{freeN8}, while \p{genN8} is ${\cal N}{=}8$ superconformal
at any $\gamma{\,}'\neq 0\,$.

One can also check that each of the two FI terms, \p{FI1} and \p{FIpsi},
is invariant under ${\cal N}{=}8$ supersymmetry \p{N82}, \p{N8V}. However,
they both break
superconformal symmetry and $SO(7)\,$.
The same symmetry properties are exhibited by the on-shell potential terms
arising
upon eliminating the auxiliary fields $A^{ik}, a^m$ and $\bar a^m$
from the sum of \p{genN8}
and \p{FI1}, \p{FIpsi}.

\section{Component actions}
Using the explicit component expansions \p{WZV1} and \p{PsiConstr},
it is straightforward to find
the component form of the superfield actions \p{free_v} and \p{conf_v}
\bea
S^{({\cal N}{=}8)}_{free} = \int dt\left[\dot x \dot x
-\frac{i}{4}\left(\omega^i\dot{\bar{\omega}}_i -
\bar\omega^i\dot{\omega}_i\right) + i\psi^{iA}\dot{\psi}_{iA}
+\frac{1}{2}A^{ik}A_{ik}
+ a^A\bar a_A\right] \equiv \int dt\, {\cal L}^{(N=8)}_{free}\,, \label{freeComp}
\eea
\bea
S^{({\cal N}{=}8)}_{sc} = \int dt \left[ x {\cal L}^{(N=8)}_{free}
+\frac{i}{2}A^{ik}\left(\psi^A_i\psi_{kA}
+ \frac{1}{2}\omega_i\bar\omega_k\right)
- \frac{1}{2}\left(\bar\omega{}^k\psi_{kA}\, a^A
+ \omega^k\psi_{k}^{A}\,\bar a_A \right)  \right].
\label{confComp}
\eea
To interpret \p{confComp} as the superconformal action, one needs to make
the shift $x = 1 + \tilde{x}\,$.
We observe that the general off-shell action \p{genN8} is specified by the linear
function $f(x) = \gamma + \gamma{\,}' x\,,$ and its derivative $f_x = \gamma{\,}'$,
in agreement
with the result of \cite{Topp} where the component
off-shell action of the multiplet ${\bf (1, 8, 7)}$ was constructed
by a different method.
Here we reproduced this result in a manifestly ${\cal N}{=}4$ supersymmetric
off-shell superfield
formalism. We took advantage of the latter to show the $F(4)$ superconformal
invariance
of the action \p{confComp}, as well as of a sum of \p{freeComp} and
\p{confComp} (with respect to modified $F(4)$ transformations).
Also, we showed that the FI terms \p{FI1}, \p{FIpsi} preserve ${\cal N}{=}8$
supersymmetry (although they break $F(4)\,$). The component
expressions for \p{FI1}, \p{FIpsi}
can be easily found
\be
S^{(v)}_{FI} = - \int dt\,c^{ik}A_{ik}\,, \quad
S^{(\psi)}_{FI} = \int dt (\xi_A\bar a^A - \bar\xi^Aa_A)\,. \label{FIcomp}
\ee
After eliminating the auxiliary fields $A^{ik}, a^A, \bar a_A$ in the sum
$S_{gen}^{(N=8)}+ S^{(v)}_{FI} + S^{(\psi)}_{FI}$ by their algebraic equations of motion,
there appears a scalar potential $\sim (\gamma + \gamma{\,}' x)^{-1}$
(plus some accompanying
fermionic terms) which is not conformal. Perhaps, conformal potentials
could be generated by coupling the multiplet ${\bf (1, 8, 7)}$ to some
additional ${\cal N}{=}8$ multiplets.

For completeness, we present the component form of the transformations \p{N81}
\bea
&& \delta_\eta x = -\eta_{iA}\psi^{iA}\,, \quad \delta_\eta \omega^i
= -2\eta^{iA}a_A\,, \quad
\delta_\eta \bar\omega^i = -2\eta^{iA}\bar a_A\,, \quad \delta_\eta A^{ik}
= 2\eta^{(i A}\dot{\psi}^{k)}_A\,, \nn
&& \delta_\eta \psi^{i A}=i\eta^A_kA^{ki} - i \eta^{iA}\dot{x}\,, \quad \delta_\eta a^A
= i\eta^{iA}\dot{\omega}_i\,, \quad \delta_\eta \bar a^A
= i\eta^{iA}\dot{\bar\omega}_i\,. \label{CompHid}
\eea
It is straightforward to check the invariance of \p{freeComp} and \p{confComp} under
these transformations.
Let us also give the $R$-symmetry transformations
belonging to the coset $SO(7)/[SU(2)]^3\,$, where $[SU(2)]^3$ is the product
of three $SU(2)$ symmetries: $SU(2)_{PG}$ acting on the indices $A$
and commuting with the manifest
${\cal N}{=}4$ supersymmetry, manifest $R$-symmetry $SU(2)_R$ acting
on the doublet indices $i, k, \ldots\,$, and
one more hidden $SU(2)_{R'}$ which rotates $\theta_i$ through
$\bar\theta_i$, $D^i$ through
$\bar D^i$, $\omega^i$ through $\bar\omega^i\,$ and $a^A$ through $\bar a^A$.
These transformations read
\bea
&& \delta_\lambda A^{ik} = \lambda^{(ik)B} a_B
- \bar{\lambda}^{(ik)}_{\;\;B}\bar a^B\,, \;
\delta_\lambda a^B = -\bar{\lambda}^{(ik) B}A_{ik}\,, \;\delta_\lambda \bar a^B
= \lambda^{(ik) B}A_{ik}\, \nn
&& \delta_\lambda \psi^{iA} = -\frac{i}{2}\left[\lambda^{(ik)A}\omega_k
+ \bar\lambda^{(ik)A}\bar\omega_k\right], \;\delta_\lambda \omega^i
= -2i\,\bar\lambda^{(ik)A}\psi_{k A}\,,\;\delta_\lambda \bar\omega^i =
2i\,\lambda^{(ik)A}\psi_{k A}\,. \label{SO7}
\eea
Here $\lambda^{(ik)B}$, $\bar\lambda_{(ik)B}$ are 6 complex parameters, which,
together with
9 real parameters of $[SU(2)]^3$, are the 21 real parameters of the group $SO(7)\,$.
The bosonic field $x$ is an $SO(7)$ singlet.

\setcounter{equation}{0}
\section{Conclusions}
In this article, using the manifestly ${\cal N}{=}4$ supersymmetric language
of the ${\cal N}{=}4, d{=}1$
harmonic superspace, we constructed a new ${\cal N}{=}8$ superconformal model
associated with the
off-shell multiplet ${\bf (1,8,7)}$ and showed that the corresponding
${\cal N}{=}8$ superconformal
group is the exceptional supergroup $F(4)\,$. We also found that the generic
sigma-model type
off-shell action of this multiplet is given by a sum of the superconformal action
which is trilinear in
the involved ${\cal N}{=}4$ superfields and the free bilinear action. The
generic action is also
superconformal, but with respect to modified $F(4)$ transformations. The
component action is in agreement
with the one derived in \cite{Topp}. The ${\cal N}{=}8$ supersymmetric
potential terms can be generated
by two superfield FI terms which break both superconformal and $SO(7)$
symmetries. An interesting
problem for further study is to see whether superconformal potentials
could be generated
by coupling this model to some other known ${\cal N}{=}8$ multiplets.
Also it would be of interest to find
out possible implications of this new superconformal model in the brane,
black holes and AdS$_2$/CFT$_1$ domains, e.g. along the lines of
refs. \cite{nscm}-\cite{GTow}, \cite{AGK}-\cite{Tbranes}.

\section*{Acknowledgements}
The work of E.I. was supported in part by the RFBR grant 06-02-16684,
the RFBR-DFG grant 06-02-04012-a, the grant DFG, project 436 RUS 113/669/0-3,
the grant INTAS 05-7928 and a grant of Heisenberg-Landau program.
He thanks Laboratoire de Physique, UMR5672 of CNRS and ENS Lyon, for
the warm hospitality extended to him during the course of this work.

\end{document}